\documentclass[12pt,a4paper]{article}

\usepackage{SGpreprint}
\usepackage{url,lastpage}

\nologohead{Adrian M. Seufert, Frank Schweitzer: \\
Aggregate Dynamics in an Evolutionary Network Model\\
\textsl{International Journal of Modern Physics C}, vol. \textbf{18}, no. 10
(2007)
}

\usepackage{epsfig}
\usepackage{amsmath}
\usepackage[numbers]{natbib}
\usepackage{pstricks}
\usepackage{pst-node,pst-plot,pst-coil}

\newcommand{\mean}[1]{\left\langle #1 \right\rangle}

\begin{document}

\begin{center}

   \textbf{\Large Aggregate Dynamics \\ 
in an Evolutionary Network Model}\\[5mm]
   \textbf{\large Adrian M. Seufert$^{1}$, Frank Schweitzer$^{2}$}\\

   \begin{quote}
     \begin{itemize}
     \item[$1$] Institute for Theoretical Physics, Technical University
       of Berlin, Hardenberg Str. 36, D-10623 Berlin, Germany

     \item[$2$] Chair of Systems Design, ETH Zurich, Kreuzplatz 5,
       CH-8032 Zurich, Switzerland, Corresponding author:
       \texttt{fschweitzer@ethz.ch}
     \end{itemize}
   \end{quote}
\end{center}

\begin{abstract}

  We analyze a model of interacting agents (e.g. prebiotic chemical
  species) which are represended by nodes of a network, whereas their
  interactions are mapped onto directed links between these nodes.  On a
  fast time scale, each agent follows an eigendynamics based on
  catalytic support from other nodes, whereas on a much slower time scale
  the network evolves through selection and mutation of its
  nodes-agent. In the first part of the paper, we explain the dynamics
  of the model by means of characteristic snapshots of the network
  evolution and confirm earlier findings on crashes an recoveries in the
  network structure. In the second part, we focus on the aggregate
  behavior of the network dynamics. We show that the disruptions in the
  network structure are smoothed out, so that the average evolution can
  be described by a growth regime followed by a saturation regime,
  without an initial random regime. For the saturation regime, we obtain
  a logarithmic scaling between the average connectivity per node
  $\mean{l}_{s}$ and a parameter $m$, describing the average incoming
  connectivity, which is independent of the system size $N$.

\textbf{Keywords:} {directed network, evolution, scaling laws} \\

PACS Nos.: 
89.75.Fb, 
89.75.Da, 
87.23.Kg, 
64.60.Cn, 
05.65.+b 

\end{abstract}

\section{Introduction}
\label{intro}

Many evolutionary processes in physical, biological or economic systems
involve elements of selfreproduction and catalytic interactions. In his
pioneering work, \citet{eigen71} pointet out their relevance for the
prebiotic evolution of macromolecules, which lead to the theory of the
\emph{hypercycle} (see also \citep{eigen-schuster-79}). The hypercycle
can be seen as a paragon of a network of \emph{cooperating agents}
\citep{hofbauer84:_evolut_system} (e.g. chemical or biological species),
which counterbalances the effect of aggressive self-replication. While
the latter one just leads to the survival of only one species --
``survival of the fittest'' -- the dependence on catalytic interaction
with other species also ensures the survival of the others and, hence, a
coexistence of agents with very different ``fitness'' levels.

Recently, the hypercycle concept has been investigated in a modified
setting, which combines the original idea of catalytic interactions with
an external dynamics of the network representing the interaction
structure. Inspired by earlier work \citep{farmer86:_autoc_replic_polym,
  waechtershaeuser90:_evolut_first_metab_cycle}
\citet{jain98:_autoc_sets_growt_compl_evolut} have focussed on the
emergence of so-called \emph{autocatalytic sets} (ACS) among agents,
which do not self-replicate individually, but only replicate by means of
the help of others. An ACS is then a cooperative structure, where
different agents interact in such a way that the links representing these
interactions form a closed cycle in terms of the network structure. Once
an ACS appears, it boosts the replication of the agents involved, which
leads to a larger growth or ``output'' of those agents involved in the
ACS. It further allows other agents not directly part of the ACS but only
linked to it, to still benefit from it as freeloaders.

Because such a catalytic replication dynamics eventually leads to a
stationary state, \citet{jain98:_autoc_sets_growt_compl_evolut} have
added a disturbance of the interaction network in terms of a so-called
``extremal dynamics''
\citep{bak93:_punct_equil_critic_simpl_model_evolut}. There, the least
performing agent, i.e. the one with the lowest output, is -- together
with its links to other agents -- replaced by a new agent that is linked
to the existing interaction network in a random way.  This network
dynamics occurs on a much slower time scale compared to the agent
dynamics itself. It ensures (i) that the dynamics of the system of agents
does not get stuck in an equilibrium state, and (ii) may allow for
``evolutionary'' scenarios towards a better performance of the whole
system.

Our work, discussed in the following, is based on the model of Jain and
Krishna (JK) described above (see also
\citep{
  jain02:_large_extin_evolut_model}).  In Chapter \ref{model} we explain
the dynamics and our numerical implementation of the JK model in more
detail. In Chapter \ref{results_cs} we reproduce some important features
of the model behavior, such as the emergence of the ACS and the crashes
and recoveries in the network structure, by means of computer simulations
that elucidate the network evolution. In Chapter \ref{Aggregate_level} we
extend our investigations to the aggregate behavior of the system, which
to our knowledge was not investigated before. In particular, we show that
the crashes and recoveries in single network realizations are smoothed
out, so that the average evolution can be described by a saturation
dynamics. We further obtain, by means of computer simulations, a
logarithmic scaling function for the average connectivity per node
dependent on the average incoming connectivity (which is a measure for
the catalytic interaction). In Chapter \ref{Conclusion} we summarize
these findings and point to further interesting extensions of the model.
In particular, we already mention the relation to recent network models
for social and economic applications.  \citep{stauffer-hohnisch06,
  holme-newman06, miskiewicz-ausloos06, koenig06, schnegg06}

\section{The Model}
\label{model}

\subsection{Population and Network Dynamics}
\label{PopNet}
The model discussed in the following was originally developed in the
context of the ``origin of life problem'': the observation that something
as structurally complex as a living cell was able to form, parting from a
random mix of chemical components in a prebiotic ``broth''
\citep{eigen71, waechtershaeuser90:_evolut_first_metab_cycle} on earth
four billion years ago.

For a modelling approach, we consider a set of $N$ prebiotic chemical
species, each of them characterized by a population $y_{i}\geq 0$
$(i=1,...,N)$. The dynamics of the variables $y_i$ shall be governed by
the following equation:
\begin{equation}
\dot{y_i}=\sum_{j}^{N} c_{ij}y_j -\phi y_i.
\label{popdyn}
\end{equation} 
$\phi$ is assumed to be a constant dilution flux, resulting e.g. from a
natural movement of raw material out of the system (say through flood or
tides). The $c_{ij}$ are the kinetic coefficients that describe the
replication of species $i$ resulting from \emph{binary} interactions with
other species $j$. For simplicity, only $c_{ij}\in\{0,1\}$ is assumed.
$c_{ij}=1$ represents a growth process of species $i$ due to the presence
of species $j$ that acts as a catalysor only.  Negative values of
$c_{ij}$ would indicate inhibitory processes that 
are neglected here. Further, self-replicating species are not allowed,
which means $c_{ii}=0$ for all $i$.

In a first approximation $\phi$ can be set to zero. This results in a
linear dynamical system of coupled first-order differential equations in
the populations $y_i$. In vector notation this reads:
\begin{equation}
\mathbf{\dot{y}}= {C}\cdot \mathbf{y}
\label{popdyn_vecnot}
\end{equation}
where ${C}$ is the matrix containing all kinetic coefficients
$c_{ij}$. The solution of the set of equations (\ref{popdyn_vecnot})
depends on the properties of the matrix ${C}$ and has the
general form: 
\begin{equation}
\mathbf{y}(t)=e^{{C}t}\mathbf{y}_0 
\end{equation} 
representing an exponential increase in time of the population vector.
To avoid the problem of exploding populations we consider the vector of
\textit{relative populations}
\begin{equation}
  \label{eq:x-i}
 x_i=\frac{y_i}{\sum_j y_j}\;; \quad \sum_j x_j = 1  
\end{equation}
Rewriting Eq. (\ref{popdyn}) by means of Eq. (\ref{eq:x-i}) gives us the
\textit{relative population dynamics}:
\begin{equation}
\dot{x_i}=\sum_{j}^{N}c_{ij}x_j - x_i \sum_{k,j}^{N}c_{kj}x_j.
\label{rel_popdyn}
\end{equation} 
Equation~(\ref{rel_popdyn}) has the property of preserving the
normalization of $\mathbf{x}$. Henceforth we will always refer to the
population vector as $\mathbf{x}$ and refer to the corresponding
population dynamics of Eq.~(\ref{rel_popdyn}). The dilution flux $\phi$
disappears in this transformation (as long as it is assumed to be equal
for every species) which gives us another reason to set it arbitrarily to
zero in Eq.~(\ref{popdyn}).

So far, we have discussed the dynamics of interacting species, where the
interaction is described by the matrix $C$ that contains the elements
$c_{ij}$ in terms of $0$ and $1$. This dynamics can be interpreted as a
\emph{catalytic network} \citep{kaneko05,lee-et97,stadler-et93,stadler91}
where the different species $i$ are represented
by nodes, and their interaction by links between these nodes, cf.  Fig.
(\ref{example_dirnet}). More precisely,
\begin{equation}
c_{ij}=\left\{ 
\begin{array}{l} 
1 \rightarrow \mbox{species $j$ catalyzes species $i$} \\
0 \rightarrow \mbox{nothing happens}
\end{array}\right.
\label{adj}
\end{equation}
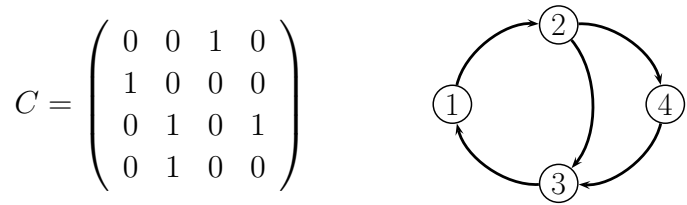
\begin{figure}[htbp]
\begin{minipage}{.45\textwidth}
\begin{displaymath}
C=\left(
\begin{array}{cccc}
0 & 0 & 1 & 0\\
1 & 0 & 0 & 0\\
0 & 1 & 0 & 1\\
0 & 1 & 0 & 0 
\end{array}
\right)
\end{displaymath}
\end{minipage}
\begin{minipage}{.45\textwidth}
\scalebox{0.7}[0.7]{
\begin{pspicture}(6,6)
 \cnodeput(0,3){1}{\Large 1}
 \cnodeput(2,4.5){2}{\Large 2}
 \cnodeput(2,1.5){3}{\Large 3}
 \cnodeput(4,3){4}{\Large 4}
 
 \ncarc[linewidth=1.5pt,arcangle=40]{->}{1}{2}
 \ncarc[linewidth=1.5pt,arcangle=40]{->}{2}{3}
 \ncarc[linewidth=1.5pt,arcangle=40]{->}{3}{1}
 \ncarc[linewidth=1.5pt,arcangle=40]{->}{2}{4}
 \ncarc[linewidth=1.5pt,arcangle=40]{->}{4}{3}

\end{pspicture}
}
\end{minipage}
\caption{A directed graph consisting of $4$ nodes (species) and $5$ links
  (interactions). Left: the corresponding interaction matrix
  \label{example_dirnet}}
\end{figure}

The matrix containing the catalytic interactions $c_{ij}$ is called the
\textit{adjacency matrix}.  The network of interactions is modeled on a
\textit{directed graph}, which means that the adjacency matrix is not
generally symmetric: $c_{ij} \not= c_{ji}$. It should be noted that the
matrix $C$ represents a linear dynamical system in Eq.~(\ref{popdyn}),
and, simultaneously a directed graph of interactions.  This means that
$C$ acquires both structural and dynamical significance.

So far, the dynamics given by Eq. (\ref{rel_popdyn}) are considered for a
fixed configuration of the matrix $C$, which translates into a fixed
network structure. In the following, we want to introduce a dynamics for
the network itself, which means an additional element in our model, where
the population dynamics is given by Eq.~(\ref{rel_popdyn}). In agreement
with \citep{jain98:_autoc_sets_growt_compl_evolut}, our main assumption
is here that the two different dynamics, the population dynamics and the
network dynamics, occur on \emph{two different time scales}. More
precisely, it is assumed that the population dynamics is fast and relaxes
into a quasi-stationary state (or attractor) soon, whereas the network
dynamics occurs on a much slower time scale, and only happens after the
population dynamics has reached its attractor. Hence, we are able to
separate these two different time scales in our computer simulations,
described below, and will measure time as ``network time'', i.e. in steps
$n$ of the network modification.

For the network dynamics, we assume that the initial network of chemical
reactions is a randomly generated graph: each $c_{ij}$ ($i\not=j$) equals
$1$ with probability $p$ and $0$ with probability $(1-p)$. Each node
contributes on average $m=p(N-1)$ links to the network, and average total
connectivity is $N\cdot m$.  The parameter $m$ is called \emph{average
  incoming connectivity}.

The rules for the network evolution are the following: 
\begin{itemize}
\item After a given time $T$ at which the population dynamics
  Eq.~(\ref{rel_popdyn}) is expected to have relaxed to its attractor
  configuration, the \textit{least fit species}, i.e., the one with the
  smallest $x_i(T)$, is determined and removed from the network along
  with all its links. If there are different species with the same
  smallest values, then from these one species is chosen at random.
\item A new species is added to the network with some small initial
  population $x_0$. The new species will take the place of the old one
  (it gets the same label), and is randomly rewired to the network with
  the same probability $p$ for establishing links that have been used in
  the initial network distribution.  Incoming and outgoing nodes are
  \textit{statistically similar}.
\item Finally, the vector of relative populations $\mathbf{x}$ is re-normalized
  with the new node.
\end{itemize}
These rules for the network evolution are intended to capture two key
features: \textit{natural selection}, in this case, the extinction of
the weakest; and the \textit{introduction of novelty}. Both of these
can be seen as lying at the heart of natural evolution. The particular
form of selection used in this model has been inspired by the ``extremal
dynamics'' of
\citet{bak93:_punct_equil_critic_simpl_model_evolut} 
In the usual setting of a \emph{mutation/selection} scheme, one has to
realize that the two parts of the dynamics act on different levels:
selection occurs on the level of the agents -- i.e. removal of the least
fittest prebiotic species, whereas mutation occurs on the level of the
agent \emph{interaction}, i.e. in terms of a random rewiring of a new
node. 

\subsection{Numerical Implementation}
\label{NumImp}

For the numerical implementation we have to deal with the two different
time scales of the model, introduced above. Here, we exploit the fact
that the population dynamics occurs on a short time scale and relaxes
fast into an attractor, whereas the network dynamics occurs on a much
slower time scale and can thus be separated. 

The key insight leading to the numerical implementation of the
\emph{population dynamics} is to see that the fixed points of the system
described by Eq.~(\ref{rel_popdyn}) correspond to the eigenvectors of the
adjacency matrix $C$. Since $C$ is a real non-negative matrix, the
\textit{Perron-Frobenius Theorem} tells us that the largest eigenvalue of
$C$ is real and positive \citep{berman94:_nonneg_matric_mathem_scien}.
Furthermore, the corresponding Perron-Frobenius eigenvector is the only
eigenvector with purely positive entries, and represents the unique
asymptotically stable attractor of the population dynamics. One way to
see this is to imagine our initial population vector $\mathbf{x}_0$ as a
linear combination of all eigenvectors of $C$. Then
Eq.~(\ref{popdyn_vecnot}) tells us that for large times $t$, the
component of $\mathbf{x}_0$ corresponding to the largest eigenvalue will
dominate all others as $\mathbf{x}(t)=e^{\lambda_1 t}\mathbf{x}_{\lambda_1}$
where $\mathbf{x}_{\lambda_1}$ is the Perron-Frobenius eigenvector.

In order to find the attractor configuration of the population dynamics,
we adopt the \textit{power method}, meaning that the vector of initial
populations $\mathbf{x_0}$ can be expanded in terms of eigenvectors of $C$:
$\mathbf{x}_0=a_1 \mathbf{x}_{(\lambda_1)}+\dots+ a_N \mathbf{x}_{(\lambda_N)}$
with $|\lambda_1|>|\lambda_2|\dots >|\lambda_N|$.  Then
\begin{equation}
C\mathbf{x}_0=a_1 \lambda_1 \mathbf{x}_{\lambda_1}+\dots+ a_N \lambda_N \mathbf{x}_{\lambda_N}
\end{equation}
and repeated iteration of this process yields
\begin{equation}
C^k\mathbf{x_0}=a_{1}^k \lambda_1^k \mathbf{x}_{\lambda_1}+\dots+ a_{N}^k
\lambda_N^k \mathbf{x}_{\lambda_N}
\label{eigen}
\end{equation}
It is clear that for large $k$, the largest eigenvalue will dominate all
others and thus $C^k\mathbf{x_0}$ will approach $\mathbf{x_{\lambda_1}}$. One
obvious problem persists though: if $C$ has an eigenvalue that is equal
in magnitude to $\lambda_1$ but with the opposite sign, then this method will
give us incorrect results if the number of iterations is even. To be
certain that we have reached the eigenvector corresponding to the
positive eigenvalue, we add a $N\times N$ unit matrix to $C$. This
increases all the eigenvalues of $C$ by one while leaving the
corresponding eigenvectors unchanged. In this way we can be sure that the
attractor we have reached is the correct one.

We point out that this method allows us to find the attractor
configuration of Eq.~(\ref{rel_popdyn}) directly by exploiting the
algebraic properties of the adjacency matrix $C$. This greatly reduces
the costs in computation time and resources compared to a standard
numerical method for solving systems of differential equations like the
Runge-Kutta method. A number $\sim N$ of iterations usually suffices
to get reasonably close to the attractor. The number of operations 
for reaching the attractor is thus of order $O(N^2)$.

The numerical implementation of the \emph{network dynamics} is
straightforward with respect to the rules given in Chapter
(\ref{PopNet}). At each time step $n$ on the network time scale, we have
to determine the smallest element, say $j$, from the population vector
characterizing the attractor (using standard algorithms). The respective
node is rewired at random with a linking probability $p$. This happens in
two stages: first, node $j$ attempts to link itself to all other nodes $i
\neq j$, which determines the outgoing links, secondly, the incoming
links of node $j$ are set by each node $i\neq j$ attempting to link to
$j$. The rewiring of node $j$, which involves $\sim 2 N$ steps, changes
the adjacency matrix $C$ that will then feed back to the population
dynamics. I.e. after the rewiring, we have to determine the new attractor
of the population dynamics as explained above. Since we know that the
population dynamics always converges to the unique attractor, the initial
conditions for that process do not matter, so we always choose
$\mathbf{x}_{0}$ as the starting point.

For the computer simulations discussed in the following, the time scale
of the model is given by the time scale of the network dynamics only,
i.e. abbreviated by $n$.

\section{Results of Computer Simulations}
\label{results_cs}
\subsection{Evolution of  Network Structure}
\label{sec:net}

\begin{figure}[htbp]
\centering{
\includegraphics[scale=0.2]{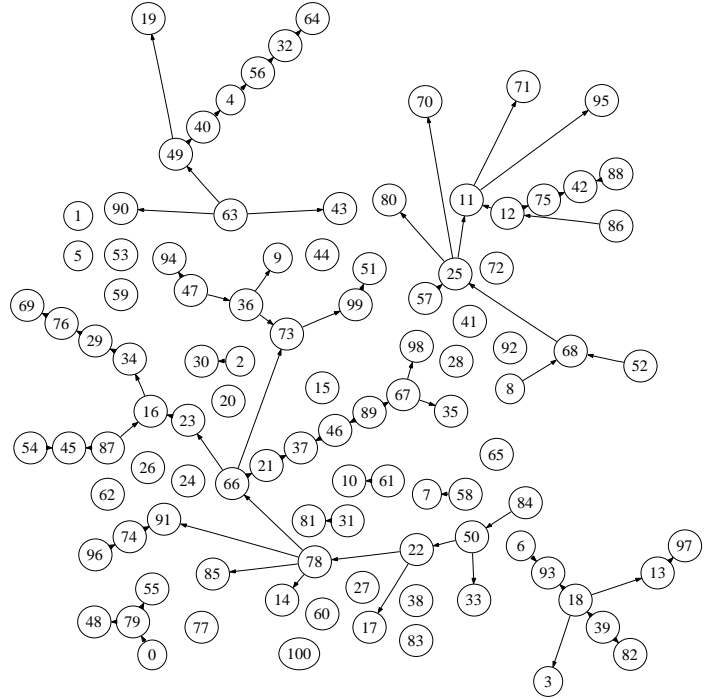}
}
\caption{Network structure in the random phase ($n=800$): several chains
  and trees exist, but no supporting structure (i.e. autocatalytic set)
  has yet emerged.  Parameters: $N=100$, $m=0.25$.  }
\label{netw-visual1}
\end{figure}

\begin{figure}[htbp]
\centering{
\includegraphics[scale=0.2]{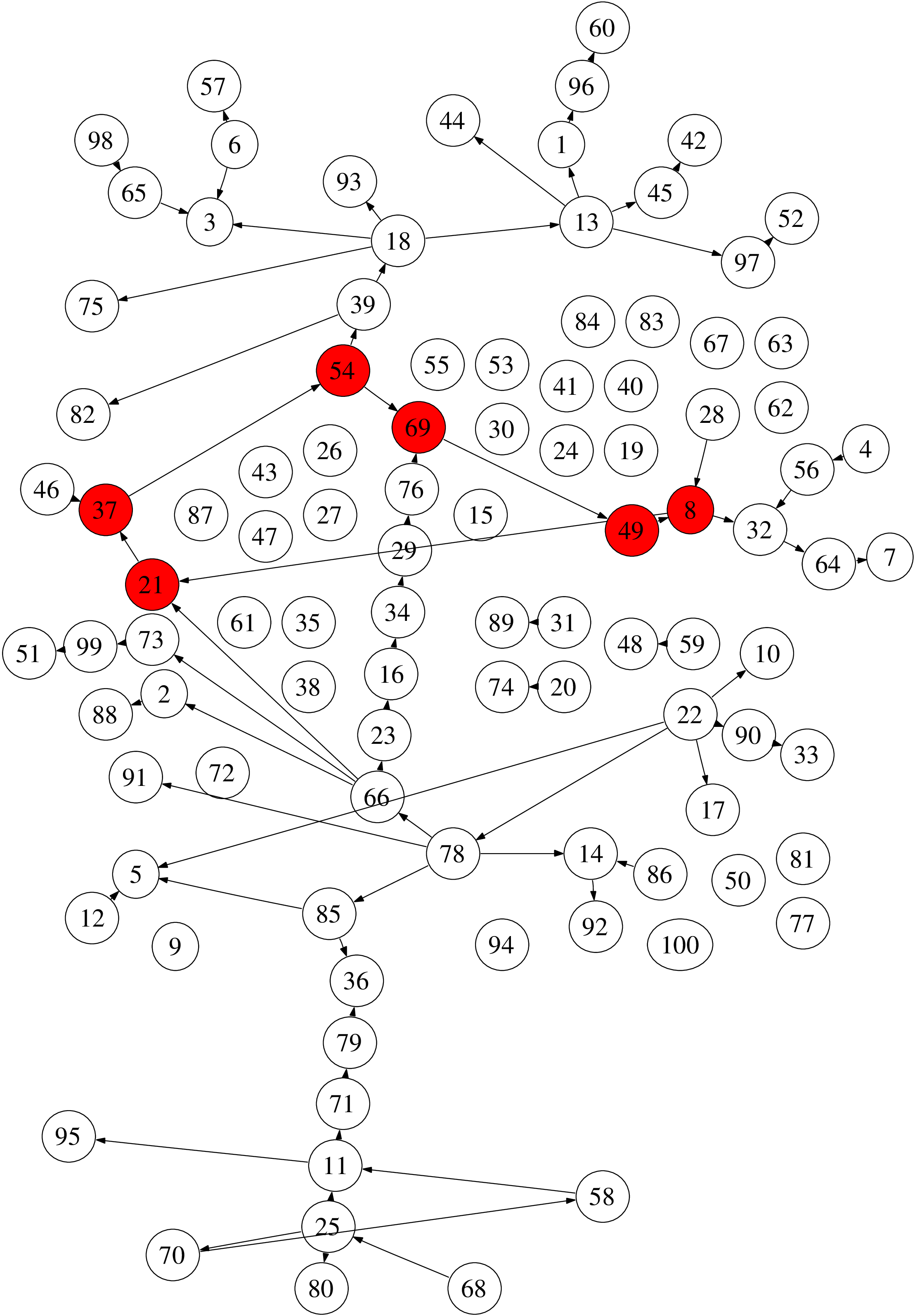}
}
\caption{Network structure after a first autocatalytic set emerges
  ($n=973$): the core of the ACS (dark nodes) consists of a six-cycle,
  which has different ``parasitic'' chains (e.g. the chain of nodes $32$,
  $64$, and $7$.  Parameters see Fig. \ref{netw-visual1}.}
\label{netw-visual2} \end{figure}

\begin{figure}[htbp] \centering{
  \includegraphics[scale=0.2]{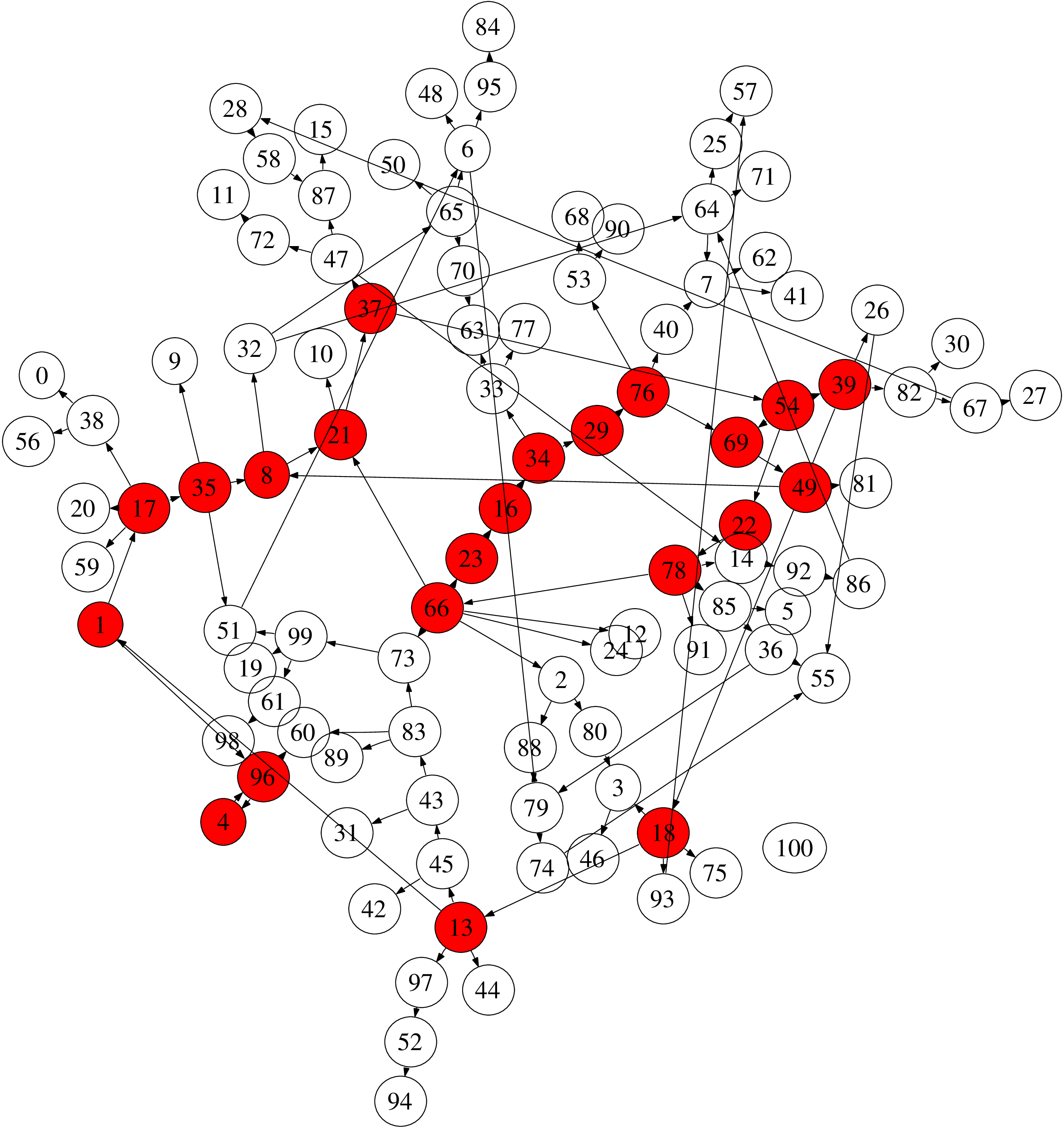} } \caption{Network
  structure after the ACS spans the whole graph ($n=1290$): the core has
  been expanded as well, but the original six-cycle still exists.
  Parameters see Fig. \ref{netw-visual1}.}  \label{netw-visual3}
\end{figure}

In Figs.~(\ref{netw-visual1})-(\ref{netw-visual3}) we show, by means of
single snapshots, three phases of development of the network according to
the dynamics described above.  In the example given, the network consists
of $100$ nodes and was generated using an average random connectivity
value $m=0.25$.  The structural properties in each of these phases will
be discussed in the following chapters.

  The first of the graphs, Fig. \ref{netw-visual1}, depicts a
typical network in an early stage of evolution.  The network is sparsely
connected and contains many singletons. Typical structures are long
chains and simple trees. The nodes located at the end of the longest
chains will usually dominate all other nodes population-wise.  In the
second graph, Fig. \ref{netw-visual2}, evolution created a new kind of
structure: a cycle consisting of six nodes (dark nodes in the picture).
Because they collectively catalyze each other, the members of the cycle
will always have non-zero population in the attractor configuration.
Thus, as long as there are singletons or separate chains and trees in the
system, members of the cycle remain "immune" to selection. This immunity
extends to all sets of nodes that have an incoming link coming from the
cycle.  These nodes form a parasitic \textit{periphery} around the
\textit{core} formed by the cycle.  Finally, the graph in
\ref{netw-visual3} shows a fully connected network.  The network is
organized around a complex core consisting of several cycles, from which
the periphery sprouts outwards. In this configuration, every node has at
least one incoming link from some other node in the network.

\subsection{The Autocatalytic Set} \label{autocat_set} The cycle of nodes
present in Fig.  (\ref{netw-visual2}) together with all the chains and
trees parting at some member-node of the cycle is an example of an
\textit{autocatalytic set} or ACS. An ACS is defined as a subgraph whose
every node has at least one incoming link from another node that belongs
to the same subgraph. The simplest ACS is a two-cycle.  The following
correspondences between ACSs and $\lambda_1$ have been found
\citep{jain98:_autoc_sets_growt_compl_evolut}:
\def\labelenumi{(\roman{enumi})} \begin{enumerate} \item An ACS always
  contains a cycle.  \item If a graph has no ACS then $\lambda_1=0$ for
  that graph.  \item If a graph has an ACS then $\lambda_1\geq 1$.  \item
  If $\lambda_1 \geq 1$, then the subgraph spanned by nodes $i$ for which
  $\mathbf{x}_{(\lambda_1),i}>0$ is an ACS.  \end{enumerate} Cycles and
structures of interlocking cycles represent \textit{irreducible}
subgraphs.  The Perron-Frobenius eigenvalue of a network is equal to the
Perron-Frobenius eigenvalue of its dominant irreducible subgraph. The
irreducible subgraph of a network that gives rise to the largest
$\lambda_1$ is called the core of the network (or alternatively the core
of the dominant ACS).  The importance of the ACS in this system is the
following: once this structure appears, all the member nodes of the ACS
have non-vanishing populations in the attractor. Because of that, they
dominate all other nodes that are outside of the ACS. The ACS is thus
robust against random mutations of its members. A small ACS appearing
randomly in this system acts like a seed for the emerging network. It
"attracts" the other nodes, since these can survive only by becoming
members of the ACS. Finally, the fully developed network at the end of
the evolution consists of a single, giant ACS.

\subsection{Network
  Connectivity} \label{NetCon} 

In order to quantitatively characterize
the network evolution, the average connectivity per node in the network
is a useful observable.  We denote this observable by $\mean{l}=l/N$,
where $l$ is the total number of links in the network.  Typically we will
choose $m$ to be in the range $(0,1)$, so in the early phase of network
evolution the graph is very sparse. For example, a network consisting of
$100$ nodes and $m=0.25$ will have on average $2\cdot m=50$ links.

\begin{figure}[h]\hspace{-5mm}
\centering{
  {\scalebox{0.5}{\includegraphics[angle=-90]{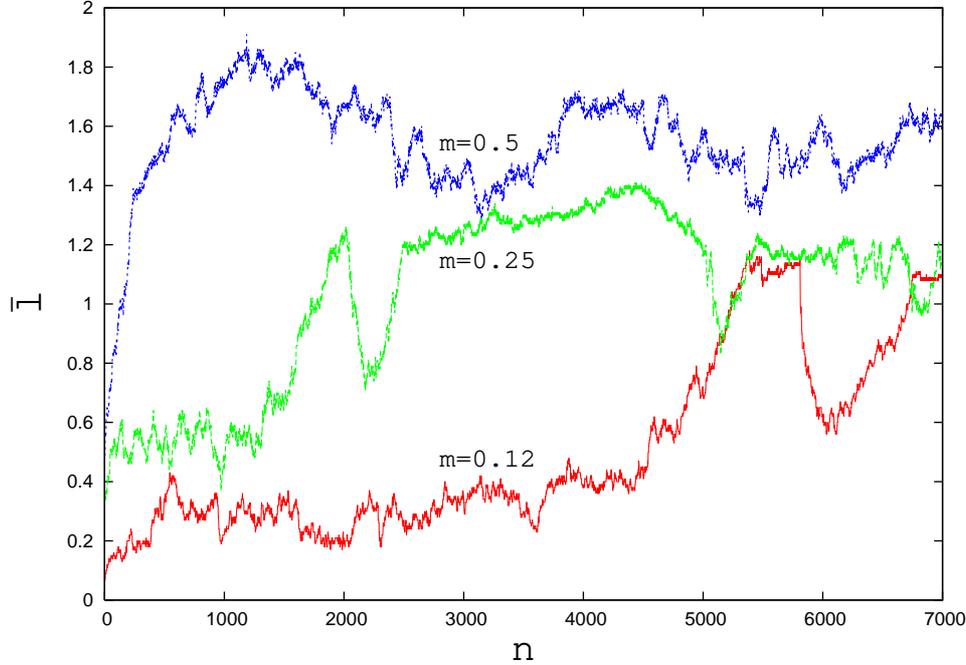}}}
}
\caption{Average number of links, $\mean{l}$ evolving over network time,
  $n$, for three values of the parameter $m$. System size $N=100$
  \label{fig1_1}}
\end{figure}

Fig.
\ref{fig1_1} shows three typical runs for the evolution of $\mean{l}$
over network time $n$, which confirm the findings of
\citep{jain98:_autoc_sets_growt_compl_evolut}(Fig. 1). During each
time-step, the network undergoes a transformation due to the selection
and random re-wiring of a node.  In effect, we have a different network
at every step, and we count and plot the total number of links in each
network divided by the number of nodes.  The runs exhibits three distinct
phases. The graphs in Figs.  \ref{netw-visual1}-\ref{netw-visual3} were
chosen to exemplify each of these phases. In the first phase (random
phase) the number of links hovers around the average expected number of
$2N\cdot m$.  This phase is followed by one of exponential increase in
the number of links (exponential phase). Finally, the number of links
stabilizes at a much higher level (steady phase).  A fundamental
structural change in the network occurs between the random phase and the
exponential phase: the emergence of an ACS.  Once it appears in the
network, it remains there until one-by-one all remaining nodes link
themselves into the ACS and it eventually spans the whole graph. Only
then will members of the ACS themselves be eligible for selection and
mutation.  The steady phase is reached when all the nodes in the network
are members of the ACS. It is characterized by the fact that the mutating
node has, on average, the same total number of links (namely $2m$) as its
replacement.  Once the ACS engulfs the whole set of nodes, its members
become eligible for selection. When a \textit{keystone species}, i.e. a
species critical to the support of the system, happens to be selected
\citep{jain98:_autoc_sets_growt_compl_evolut}, the network loses its
supporting structure and larger \textit{crashes} occur. In
Fig.~\ref{fig1_1} that is the case at $n\approx 5800$ for the $m=0.12$
run, and at $n\approx 2000$ and $n\approx 5000$ for the $m=0.25$ run. In
longer runs, the number of crashes of this sort can be quite substantial,
including crashes that completely destroy the ACS forcing the system to
"start over".  It is clear that for any $m>0$ an ACS will always emerge
eventually. The interesting point is that, although we start from a
random graph and introduce random mutations, the network resulting from
the graph-spanning ACS is highly non-random. We repeat here that the
probability of a graph with $N$ nodes and an average of $m$ links per
node being an ACS is given by:\citep{jain02:_large_extin_evolut_model}
\begin{equation}
  P=\left[1-\left(1-\left[\frac{m}{\left(N-1\right)}\right]
    \right)^{N-1}\right]^N \end{equation} which declines exponentially
with $N$ when $m\sim O(1)$.

\section{Network Structure at the Aggregate
  Level} \label{Aggregate_level} 
\begin{figure}[h!] \hspace{-5mm}
\begin{minipage}{0.92\linewidth} 
\begin{minipage}[c]{0.44\linewidth}
  \rotatebox{-90}{\includegraphics[width=0.9\textwidth]{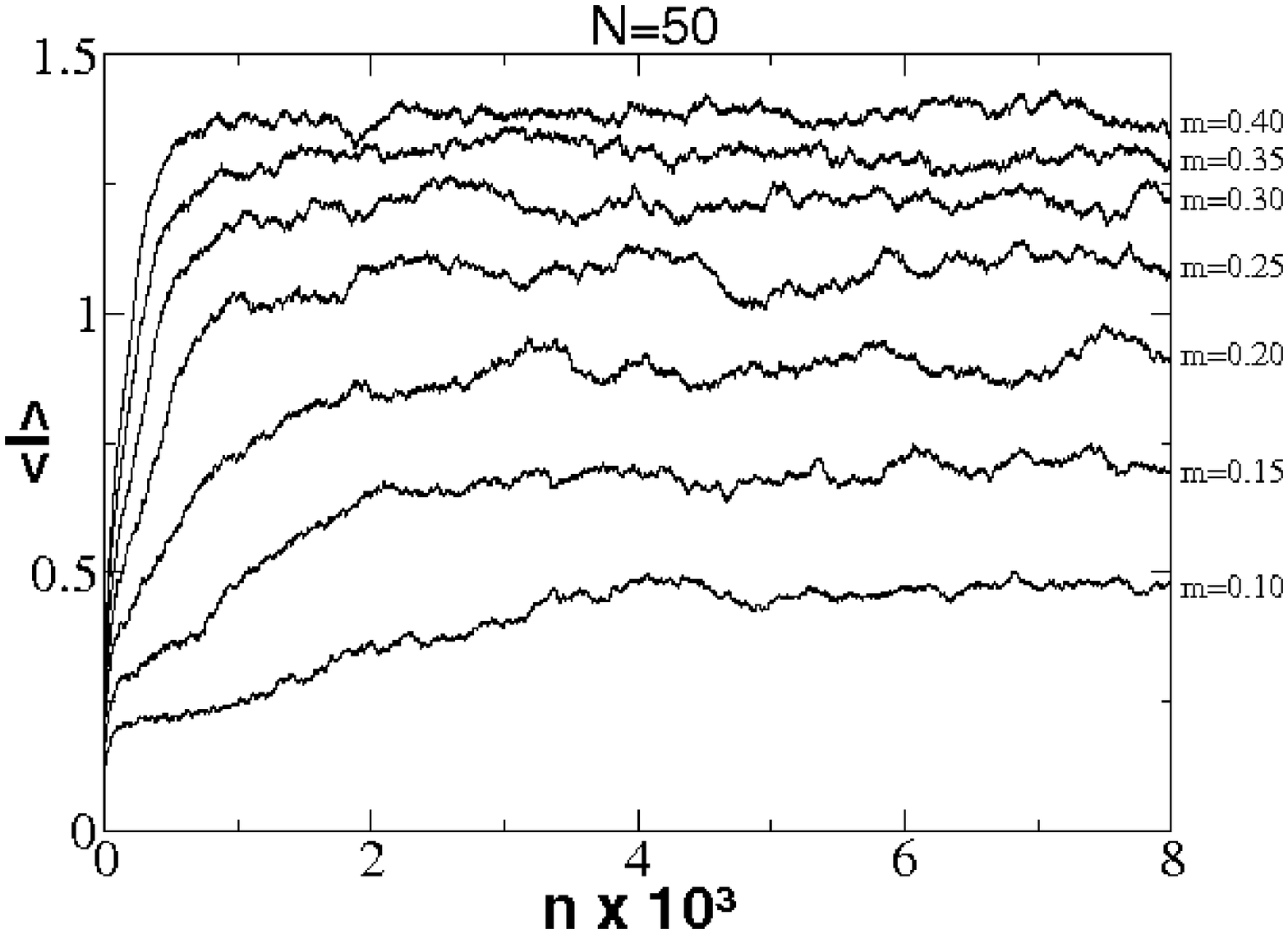}}
\end{minipage}\hfill 
\begin{minipage}[c]{0.44\linewidth}
  \rotatebox{-90}{\includegraphics[width=0.9\textwidth]{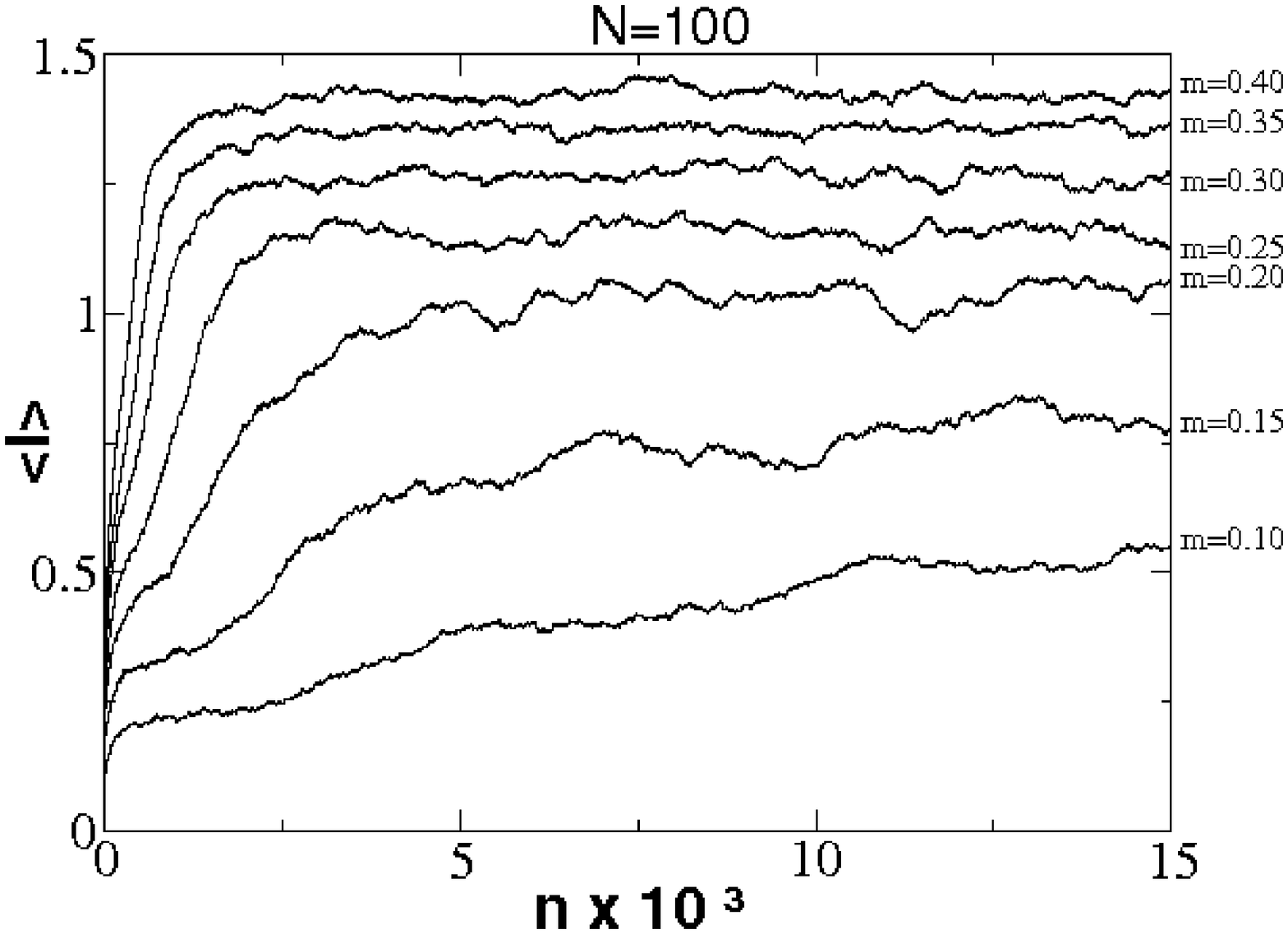}}
\end{minipage}

\begin{minipage}[c]{0.44\linewidth}
  \rotatebox{-90}{\includegraphics[width=0.9\textwidth]{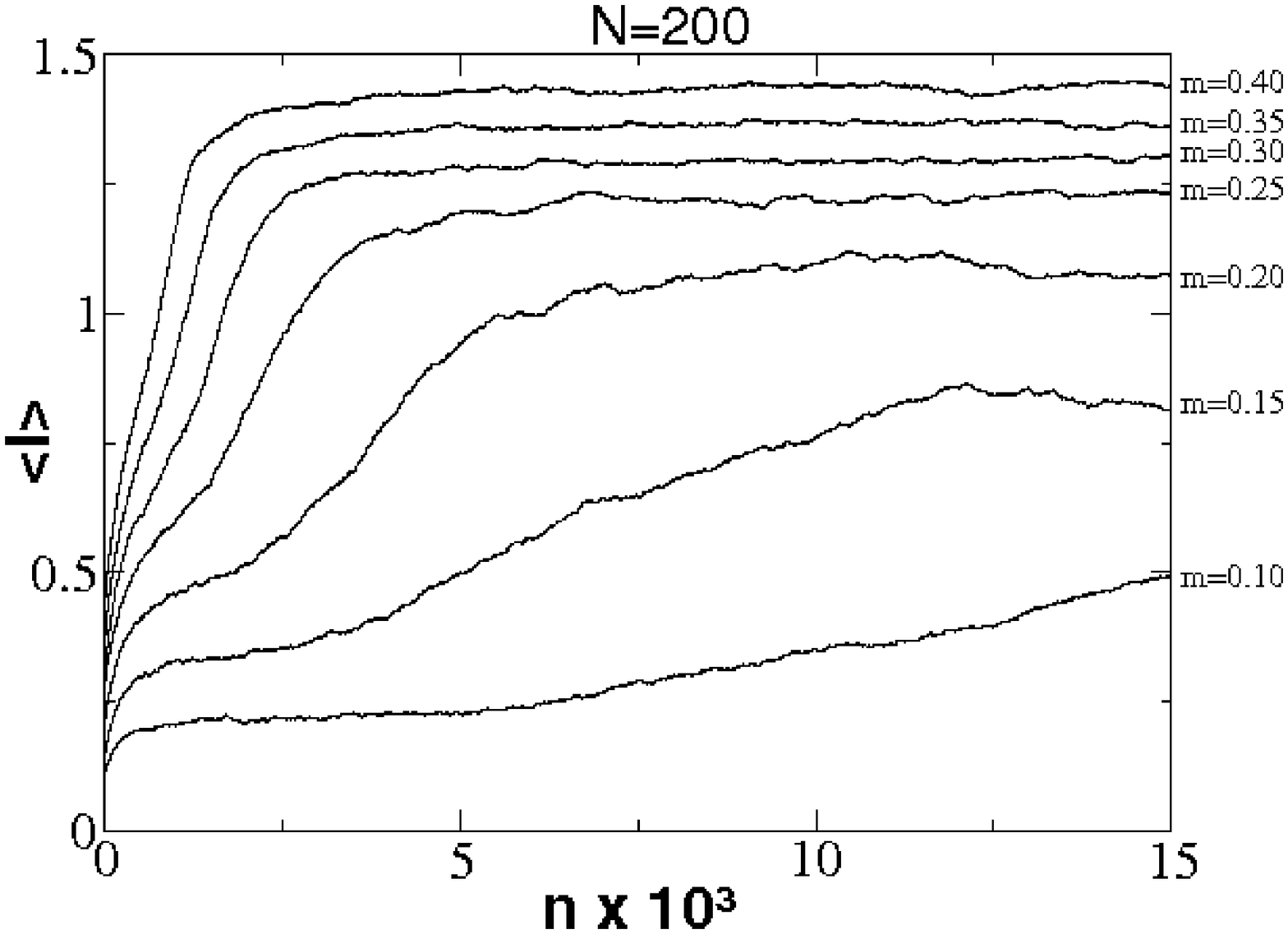}}
\end{minipage} \hfill 
\begin{minipage}[c]{0.44\linewidth}
  \rotatebox{-90}{\includegraphics[width=0.9\textwidth]{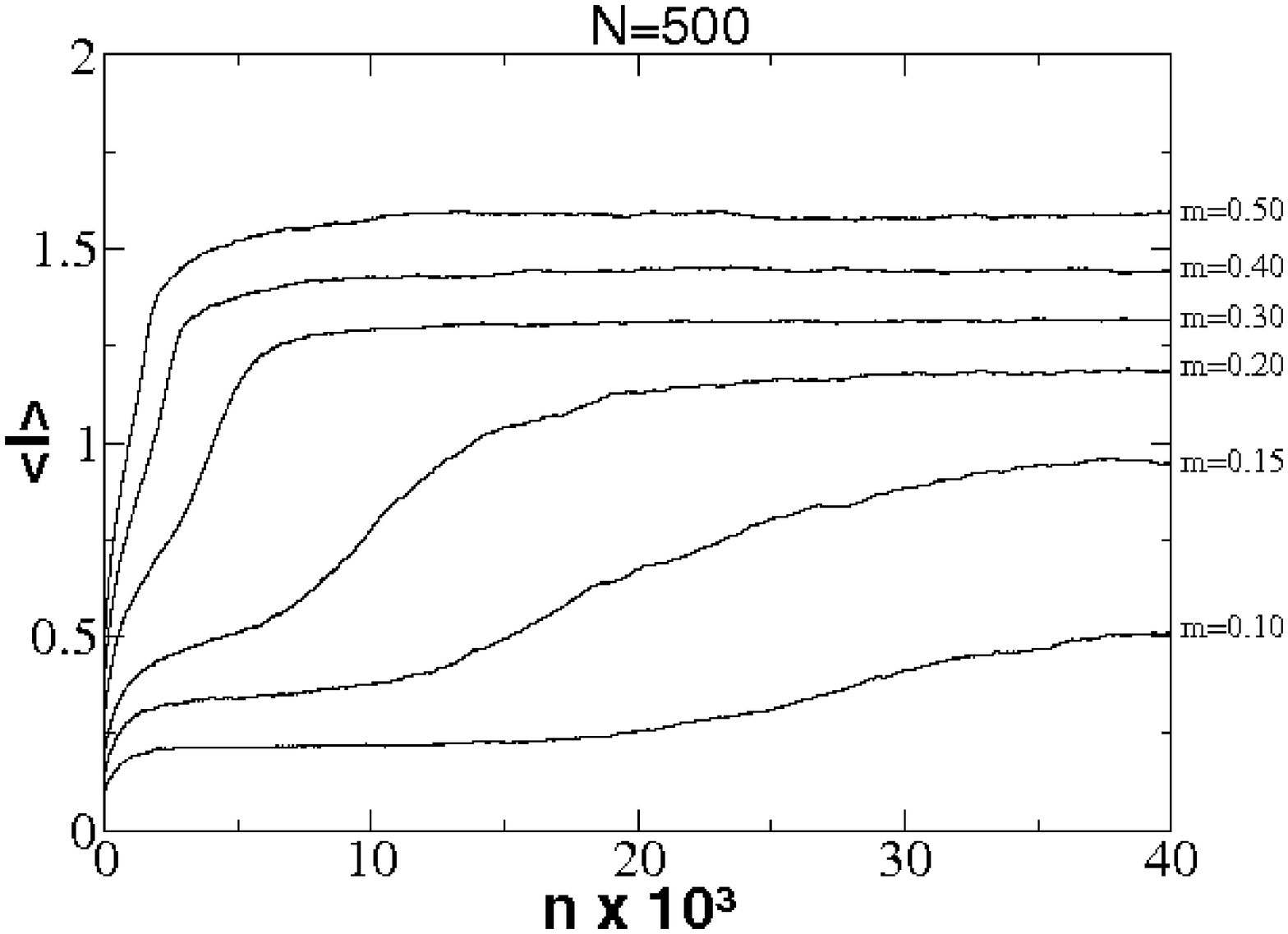}}
\end{minipage}
\end{minipage}
\caption{Average connectivity 
  per node $\mean{l}$ after averaging over $100$ runs. The
  different graphs represent different system sizes. top left: $N=50$.
  top right: $N=100$. bottom left: $N=200$.  bottom right: $N=500$.
  Evolution time is chosen according to system size.
  \label{ensemble_links}}
\end{figure}

We have seen in Chapter \ref{results_cs}
that the typical simulation run in this model exhibits three distinct
phases (Fig.~\ref{fig1_1}): the random phase, the growth phase, and the
saturation phase.  Over long time scales, these patterns tend to repeat
themselves, as fully developed networks (that is, graph-spanning ACSs)
undergo core-shifts and other transformations
\citep{jain02:_large_extin_evolut_model} that destroy the network
supporting structure. This means that, despite these characterizations,
the long term behavior of a \emph{single} run is inherently random: the
next big crash remains impossible to predict.  We will refer to these
individual runs as \textit{``microscopic''} realizations.  Opposed to
that, in the following we will look at this system from a ``macroscopic''
point of view.  This means that we will be interested in the
\emph{aggregate behavior} of a system characterized by its parameter $m$
when the simulation is repeated multiple times. Single runs ranging from
$8\cdot 10^3$ to $10^6$ time steps have been created, and ensembles of
$100$ runs analyzed. Each run starts from different initial conditions
and random numbers. Thus, the computational resources used in these
simulations were considerable, making simulations employing more than
$500$ nodes too time consuming to be realized. The data points for every
instant in simulation time are averaged over the 100 microscopic runs.
As we vary the system size $N$, the linking probability $p$ is rescaled
inversely to $N$ so as to keep the average incoming connectivity
$m=p\cdot (N-1)$ constant. Fig. \ref{ensemble_links} depicts a set of
macroscopic runs for different $m$ values and varying system size.

To save computing time and resources, we adjusted the
network evolution time $n$ to our needs in each case. Thus the $N=50$
system evolves over $8000$ time steps, while the $N=500$ needs $40000$
time steps to develop.  To summarize our findings for the aggregate
level, (i) we see that all of the curves are smooth and that none exhibit
the kind of abrupt break-in that we observe in the microscopic runs.
This represents the average evolutionary process of this system. That is,
after the period of exponential growth of the average connectivity per
node, the systems settle into a statistically stable condition of high
average connectivity. These observations are valid for all of the studied
system sizes and all values of $m$.  We emphasize, (ii) that the
transient is markedly different in the macroscopic run.  For $m>0.15$ the
initial random phase is almost in-existent in all systems. Instead, they
enter the growth phase almost immediately, reaching saturation faster for
larger $m$. The random phase is visible for smaller values of $m$ only.
Eventually, we note (iii), that for large $m$ the \emph{saturation value}
of $\mean{l}$ is independent of system size $N$.  That is, this system
scales well with the number of nodes. In the limit \begin{displaymath}
  \lim \left\{ \begin{array}{ll}
      \scriptstyle N\rightarrow \infty \\
      \scriptstyle p\rightarrow 0 \end{array} \right\} p \cdot (N-1) =
  \mbox{const}=m \end{displaymath} $\mean{l}$ is dependent only on $m$ as
we will see below.  In Fig. \ref{ensemble_links_saturation} we plotted
the averaged values of $\mean{l}$ in the saturated regime as a function
of $m$. We call this new variable $\mean{l}_{s}$.  The values used here
were computed by taking the average of $\mean{l}$ from every curve in
Fig.~(\ref{ensemble_links}) starting from a point at which saturation was
deemed to have been reached. 
We then used a least squares method to fit a function to the data points
that best approaches the qualitative shape of the data. As one can see
from the plots, we find that $\mean{l}_{s}$ is a slow, monotonically
increasing function of $m$, that can be well approximated by a function
of the form \begin{equation} \mean{l}_{s}(m)=a\cdot \ln(m) + b
  \label{logsc} \end{equation} with the constants \begin{equation*}
  a=2.06 \pm 0.021 \;;\quad b=0.66 \pm 0.042 \end{equation*} I.e., as a
new finding observed on the aggregate level, we obtain a
\emph{logarithmic scaling} of the saturated average connectivity per
node, $\mean{l}_{s}$ with $m$. Again, this scaling is independent of
system size $N$: both coefficients $a$ and $b$ vary only very slightly
with $N$.  \begin{figure}[h!] \hspace{-5mm} \begin{minipage}{\linewidth}
    \centering{
      \rotatebox{-90}{\scalebox{0.28}{\includegraphics{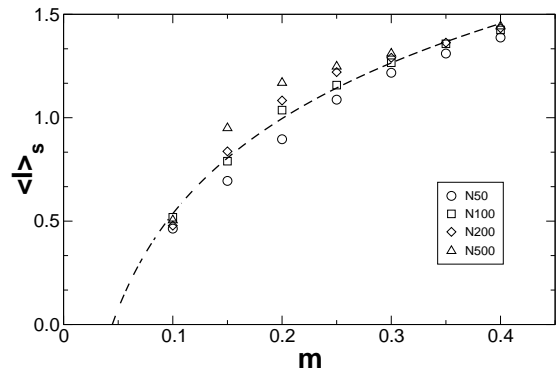}}}
    } \end{minipage} \caption[]{Saturated average connectivity per node
    $\mean{l}_{s}$ dependent on average incoming connectivity $m$.
    System sizes are the same as in Fig.~\ref{ensemble_links}. The plots
    are fitted by the logarithmic scaling of Eq. (\ref{logsc}).
    \label{ensemble_links_saturation}} \end{figure}

\section{Summary and
  Conclusion} \label{Conclusion} In this paper we have analyzed a model
of network evolution that was recently introduced by
\citet{jain98:_autoc_sets_growt_compl_evolut} \citep{
  jain02:_large_extin_evolut_model} as a combination of a hypercycle
dynamics for the nodes and an external network dynamics for the links
representing the catalytic interactions between the nodes.  The basic
concepts of the model are presented in Chapter \ref{model}. The
population dynamics is described by Eq.(\ref{eq:x-i}), while the directed
network of the species that catalyze each other are considered in the
adjacency matrix $C$ in Eq. (\ref{adj}). The evolution of the network is
then governed by the natural selection of the weakest species and the
introduction of novelty by new nodes. In such way a coupling of
population dynamics with network dynamics is realized.  In the first part
of the paper, we have investigated the evolution of the network by means
of single runs.  The numerical implementation of the model is presented
in some detail in Chapter \ref{NumImp}.  The crucial part of the
Perron-Frobenius eigenvectors of the adjacency matrix $C$ in finding the
attractor population for every network update step is emphasized. This
option is used in the power method based on Eq. (\ref{eigen}). In an
iterative procedure the weakest species of the attractor is eliminated
and replaced with a newly linked node.  As a result of these computer
simulations, Figs. \ref{netw-visual1}-\ref{netw-visual3} elucidate the
structural changes of the network by means of different snapshots. They
show typical cases of the phases of development of the network.  Starting
from an initial random phase the network processes to highly structured
configurations. Complex structures evolve that can be characterized by
Autocatalytic Sets (ACS). Our simulations validate the spontaneous
occurrence and time evolution the ACSs described in
\citep{jain98:_autoc_sets_growt_compl_evolut}.  Additionally we have
computed the total number of links $l$ over network evolution time $n$
for different incoming connectivities $m$.  In agreement with the
findings in
\citep{jain98:_autoc_sets_growt_compl_evolut,jain02:_large_extin_evolut_model}
our simulations in Fig. \ref{fig1_1} show an increase of $l$ for higher
values of $m$ as well as crashes and recoveries of the evolving ACSs.
The main new results are presented in Chapter \ref{Aggregate_level},
where the aggregate network dynamics is investigated.  We have raised the
question how the average connectivity per node $\mean{l}$, is influenced
by varying values of the incoming connectivity $m$ and therefore analyzed
multiple simulation runs.  First we observed no abrupt break-ins in the
average connectivity $\mean{l}$. This observance is opposite to the
crashes we have seen for the supporting structures, the ACSs, during
network evolution. After an exponential growth $\mean{l}$ saturates into
a stable condition of high average connectivity. Second the saturation is
reached for higher values of $m$. And third we have found a logarithmic
scaling of $\mean{l}$ in Eq.  (\ref{logsc}) on $m$ for all system sizes
$N$. Furthermore the saturation value of $\mean{l}$ is independent of
$N$.  There exist different ways to extend the basic model discussed in
this paper. In \citep{seufert2} we have studied the impact of a selection
mechanism on the performance of the system and its network structure by
introducing a selection temperature and a performance threshold
selection. We already found evidence for a critical value of this
selection threshold for the global performance of the system. Moreover
the threshold plays an important role in size and life span of the core
of the ACS. Other future investigations of the model may involve the
emergence of hierarchical organization resulting from the network
evolution. Hierarchies are already discussed for different network
topologies \citep{vazquez03:_growin, ravasz03:_hierar_organ_compl_networ,
  trusina04:_hierar_measur_compl_networ} and are also investigated in
directed networks \citep{variano04:_networ_dynam_modul,
  PhysRevLett.88.048701}.  The model discussed in this paper is seen as
an agent-based model, where each agent represents a prebiotic chemical
species. Agents are assumed as nodes of a network, where the links
represent the catalytic interactions between the species. Because each
node follows a deterministic eigendynamics, eq. (\ref{rel_popdyn}), the
model can be also regarded as a system of coupled differential equations.
This however ignores the fact that the links (i.e. the couplings between
the equations) change on a second time scale in a non-predictable way,
because of the removal of the least fittest node and the random rewiring
of a new node. Thus, from a methodological viewpoint, we rather address
the model as an agent-based one. This perspective may hold regardless of
the question whether the agents represent the population of a species, or
individual entities, such as single companies in an economic setting.  In
fact, the catalytic network model discussed here, despite its simplicity,
may serve as a good starting point for the study of the relationship
between network structure and dynamics in several fields of research, in
particular in the economic and social domain
\citep{stauffer-hohnisch06,miskiewicz-ausloos06,holme-newman06,schnegg06}.
For example, catalytic or hypercycle interaction can be used to model
skills in economic organizations \citep{padgett97}, or exchange of
knowledge in innovation networks \citep{koenig06}. Using some more
realistic rules for rewiring the network, one can then observe the
emergence of a number of smaller ACS \citep{koenig06} instead of just one
giant ACS in the current model. This affects the breakdown probabilities
of the network consideraby.  Cooperative networks related to the ones
obtained from this model are also present in firm interactions. It was
shown, for example, that financial systems form cooperative networks of
ownership relationships with scale free properties
\citep{Garlaschelli/al:05:SF}.

\subsection*{Acknowledgement}
The authors wish to thank Michael K{\"onig} (Zurich) for his help in
redrawing Figs. \ref{netw-visual1}-\ref{netw-visual3},
\ref{ensemble_links} and for comments on Chapters \ref{intro},
\ref{Conclusion}.

\bibliographystyle{plainnat}
\bibliography{net_arxiv}

\end{document}